# A Big Data Analytics Framework to Predict the Risk of Opioid Use Disorder

Md Mahmudul Hasan[1], Md. Noor-E-Alam[1*], Mehul Rakeshkumar Patel[1], Alicia Sasser Modestino[2,3], Leon D. Sanchez[4,5], Gary J. Young[6,7,8]

[1] Mechanical and Industrial Engineering, Northeastern University, Boston, Massachusetts, United States of America,

[2] Public Policy and Urban Affairs and Economics, Northeastern University, Boston, Massachusetts, United States of America

[3] Dukakis Center for Urban and Regional Policy, Northeastern University, Boston, Massachusetts, United States of America

[4] Harvard Medical School, Boston, Massachusetts, United States of America

[5] Department of Emergency Medicine, Beth Israel Deaconess Medical Center (BIDMC)

[6] D'Amore-McKim School of Business, Northeastern University, Boston, Massachusetts, United States of America

[7] Center for Health Policy and Healthcare Research, Northeastern University, Boston, Massachusetts, United States of America

[8] Bouvé College of Health Sciences, Northeastern University, Boston, Massachusetts, United States of America

[*]Corresponding author: Md. Noor-E-Alam, Northeastern University, School of Engineering, 360 Huntington Avenue, Boston MA 02115, email: mnalam@neu.edu, Tel.: +1-617-373-2275

**Abstract**

Overdose related to prescription opioids have reached an epidemic level in the US, creating an unprecedented national crisis. This has been exacerbated partly due to the lack of tools for physicians to help predict the risk of whether a patient will develop opioid use disorder. Little is known about how machine learning can be applied to a big-data platform to ensure an informed, sustained and judicious prescribing of opioids, in particular for commercially insured population. This study explores Massachusetts All Payer Claims Data, a de-identified healthcare dataset, and proposes a machine learning framework to examine how naïve users develop opioid use disorder. We perform several feature selections techniques to identify influential demographic and clinical features associated with opioid use disorder from a class imbalanced analytic sample. We then compare the predictive power of four well-known machine learning algorithms: Logistic Regression, Random Forest, Decision Tree, and Gradient Boosting to predict the risk of opioid use disorder. The study results show that the Random Forest model outperforms the other three algorithms while determining the features, some of which are consistent with prior clinical findings. Moreover, alongside the higher predictive accuracy, the proposed framework is capable of extracting some risk factors that will add significant knowledge to what is already known in the extant literature. We anticipate that this study will help healthcare practitioners improve the current prescribing practice of opioids and contribute to curb the increasing rate of opioid addiction and overdose.

## 1. Introduction

According to Center for Disease Control and Prevention (CDC), the opioid overdose epidemic—a national public health emergency—takes an average of 130 lives in the US every day (CDC, 2017a), more than 35% of which are attributable to the overdose of legally obtained prescription opioids (CDC, 2017b). Opioids were responsible for approximately 47,736 deaths in 2017, and opioid overdose deaths (deaths involved prescription and illegal opioids) were six times higher in 2017 than 1999 (CDC, 2017a), the largest increase in overdose-related deaths in this country's history. This statistic is difficult to fully grasp—shocking in its scope and overwhelming in the utter complexity of trying to curtail the overflow of legal and illicit drugs that have created this national crisis. Although strong evidence for long-term benefits of opioid therapy for chronic pain management is still lacking (Chapman et al., 2010), almost 58 opioid prescriptions were written for every 100 Americans with an average of 3.4 opioid prescriptions issued per patient (CDC, 2019).

Lack of stringent restrictions on the flow of prescription opioids increases the risk of developing opioid use disorder without a prescription due to the high abuse potential for opioids and risk of diversions (CDC, 2014). A key challenge for handling opioid crisis is  the difficulty in predicting whether a patient is likely to become addicted to opioids. Currently, when prescribing opioids, physicians often use risk assessment tests based on a brief risk interview. However, prior study has shown that the Opioid Risk Tool (ORT) and Screener and Opioid Assessment for Patients with Pain-Revised (SOAPP-R) inadequately predict which patients will be misusing opioid medications (Jones & Moore, 2013). Thus, opioids are being prescribed at an alarming rate and current risk prediction methods are apparently inefficient, a duo which forces the whole nation to embrace a pressing epidemic of addiction and overdose. Opioid-overdose related fatalities could be prevented if better risk assessing methods to ensure an informed opioid prescribing practice could be developed.

However, there has been little in-depth research on predicting the risk of developing opioid use disorder beforehand to help physicians prescribe them in an informed way. Several prior research only used regression based techniques to examine the association of risk factors with opioid use disorder (opioid abuse, misuse and/or dependence)  (Ciesielski et al., 2016; Dufour et al., 2014; Edlund et al., 2010; Ives et al., 2006; Rice et al., 2012; Skala et al., 2013; Thornton et al., 2018; Turk, Swanson, & Gatchel, 2008; White, Birnbaum, Schiller, Tang, & Katz, 2009) and overdose (Glanz et al., 2018). Patients' demographics (e.g., age, ethnicity, geographic regions) (Cepeda, Fife, Chow, Mastrogiovanni, & Henderson, 2013), prior clinical history of mental illness (Rice et al., 2012; Skala et al., 2013), non-opioid substance use disorder (Rice et al., 2012), tobacco abuse (Skala et al., 2013), and non-dependent alcohol abuse (McCabe, Cranford,

& Boyd, 2006; Rice et al., 2012; Skala et al., 2013) were considered as potential risk indicators for opioid use disorder. Research conducted in (Brummett et al., 2017; Liang, Goros, & Turner, 2016) used logistic regression and found that substance abuse was the greatest predictor of opioid overdose for both male and female. However, the higher odds ratio related to male indicated stronger association between opioid overdose and male patients. These results were reaffirmed by other studies (Ciesielski et al., 2016; Rice et al., 2012) using logistic regression techniques, concluding that males were more likely to experience opioid use disorder. Another study (Thornton et al., 2018) applied logistic regression and found that the initial opioid regimen is a strong predictor of chronic opioid therapy. A cox regression model (Glanz et al., 2018) was able to distinguish high-risk and low-risk patients for naloxone prescriptions with a predictive accuracy of 66-82%. Therefore, prior research (Brummett et al., 2017; Ciesielski et al., 2016; Glanz et al., 2018; Hylan et al., 2015; Liang et al., 2016; Rice et al., 2012; Thornton et al., 2018) commonly used multi-variate logistic regression and cox regression models that achieved sub-optimal predictive performance. Besides, the reported predictive accuracy may also suffer from biased estimation as no oversampling methods were adopted to take care of the class imbalanced issue (i.e. the number of patients with opioid use disorder is significantly smaller than the patients without opioid use disorder), which potentially limit the reliability and applicability of those models to predict opioid use disorder and overdose (Chawla, Bowyer, Hall, & Kegelmeyer, 2002; Japkowicz, 2000).

Several prior studies utilized data mining and machine learning techniques that perform with higher accuracy while predicting outcomes in a variety of healthcare applications (Islam, Hasan, Wang, Germack, & Alam, 2018). Such studies include but are not limited to predicting mortality of intensive care unit (ICU) patients (Le et al., 1984; Luo, Xin, Joshi, Celi, & Szolovits, 2016; Pirracchio et al., 2015; Poole et al., 2012; Rosenberg, 2002), sepsis related organ failure assessments (Gultepe et al., 2013; Ribas et al., 2011; Simpson, 2016), breast cancer survivability (Delen, Walker, & Kadam, 2005), risk of 30-day hospital readmission for patients with heart failure, pneumonia, serious mental illness (SMI) (Chin, Liu, & Roy, 2016; Kansagara et al., 2011), generating treatment plan for diseases such as ulcers (Cho, Park, Kim, Lee,

& Bates, 2013), and depressions (Klein & Modena, 2013). However, research leveraging advanced big data analytics to precisely and proactively predict the risk of developing opioid use disorder is still in its infancy. Only a few studies used machine learning models in predicting opioid dependency (Ellis, Wang, Genes, & Ma'ayan, 2019), overdose (Lo-Ciganic et al., 2019), detecting misuse by utilizing emergency medicine service (EMS) data (Prieto et al., 2020) and opioid cessation (Cox et al., 2019). However, research performed in (Lo-Ciganic et al., 2019) using claims data only considered individuals with age above 65 (Medicare beneficiaries), while Ellis et al. (2019) utilized lab tests and vital signs from electronic health record (EHR) data and only implemented a single classifier—Random Forest. Cox et al. (2019) only included 6,188 African Americans and 6,835 European Americans, leading to a narrowly defined study cohort.

Therefore, after reviewing the extant literature on the applicability of machine learning and multi-variate regression based techniques to predict risk of opioid use disorder and overdose, we identified the following limitations of previous research: (1) smaller sample sizes that did not support robust analytical files, (2) a narrow set of pre-selected clinical and socio-demographic factors for inclusion in the predictive model, (3) only multi-variate logistic regression and cox regression based predictive analytics were used, which only examined the associations between risk factors and opioid used disorder and/or overdose, and (4) reported low predictive accuracy, which reduces the reliability and applicability of existing models to predict opioid use disorder. These limitations of the existing studies create an avenue for further research to deploy data mining and machine learning techniques on large-scale healthcare claims dataset to predict opioid use disorder, and to investigate their predictive power. Hence, this study intends to provide the knowledge and tools necessary for improving opioid-prescribing practice and help curb the country's rapidly increasing rate of opioid addiction. The essential idea is to harness the potential of big-data analytics to identify patient-level factors associated with opioid use disorder, and subsequently develop a model to predict the likelihood of developing such opioid use disorder with higher accuracy. We anticipate that big-data analytics can potentially transform information into knowledge about how a naïve opioid patient progresses towards opioid use disorder, potentially supporting the development of future interventional strategies to prevent subsequent drug overdose. The results of our investigation will help physicians predict the risk of a patient to develop opioid use disorder at a later stage after being prescribed opioids.

The key contributions of this study are:

1. We develop a big data analytics framework that can analyze large-scale healthcare claims datasets in an effort to investigate patients' long-term opioid usage pattern. The proposed framework reveals how an opioid naïve patient may develop opioid use disorder at a later stage after an initial opioid prescription, and determines the influential features that are significant predictors of such opioid use disorder. Because there is a paucity of structured datasets populated with influential risk factors of opioid use disorder, the development of a reliable predictive model has been a challenging task. However, with our study, we show that the use of predictive analytics to inform judicious opioid prescribing practice is feasible with administrative claims datasets that provide information regarding patients' clinical characteristics. We successfully leverage several feature selections and a class balancing technique, supported by a large number of commercially insured population to develop a framework that has not been proposed yet to address opioid use disorder.

2. Unlike existing studies which either focused on specific subgroups of population such as Medicare individuals with age above 65 (Lo-Ciganic et al., 2019), African Americans and European Americans (Cox et al., 2019), or used a single tree-based classifier (Ellis et al., 2019), this study includes commercially insured population with age under and above 65 years in the analytic sample and implements a set of machine learning models to predict opioid use disorder. We demonstrate that, apart from Logistic Regression, other machine learning algorithms superiorly predict the likelihood of opioid use disorder using a large analytic sample that is class imbalanced, and entails skewness and sparsity of features. Furthermore, our study achieved superior predictive performance as compared to existing study. Along with higher predictive power we identify risk factors, which are significant predictors of opioid use disorder, and will add knowledge to the existing literature.

The rest of the paper is organized as follows: Section 2 presents the materials and methods used, Section 3 demonstrates the results and provides a discussion of the findings. Finally, Section 4 concludes with a summary of the work conducted in this study along with some limitations and future research directions.

## 2. Materials and Methods

### 2.1. *Data Source*

We use the *Massachusetts All Payer Claim Datasets (MA APCD)(CHIA)* which we currently house at the Northeastern Center for Health Policy and Healthcare Research, and is overseen by the Center for Health Information and Analysis (CHIA). This unique database includes all medical claims, pharmacy claims, and member eligibility information associated with commercial insurance claims in Massachusetts between 2011 and 2015. The dataset provides unique identifiers for all patients and types of providers (e.g., hospitals, physicians, nursing homes, rehabilitation facilities) that can be used to link claims for individuals across files. The pharmacy claims file contains data for approximately 470 million prescriptions and the medical claims file has approximately 1.63 billion claims for Massachusetts residents. The pharmacy claim file contains information on each prescription claim including the type of medication, the dosage, and the days' supply. The medical claim file includes information pertaining to a patient's clinical condition including principal diagnosis recorded as ICD-9 codes and ICD-10 codes for claims after October 1, 2015, services and procedures received, and payment (i.e., how much the provider received from the health plan for the services provided). Both the pharmacy and medical claims files can be linked to the member eligibility file that contains information on age, gender, insurance status (e.g., name and type of plan including co-pays and deductible).

### 2.2. *Proposed Framework for Predicting Opioid Use Disorder Using Big-data Analytics*

As previously mentioned, there is over prescriptions of opioids for acute and chronic pain management that can result in an increased risk of addiction and overdose (Fishbain, Cole, Lewis, Rosomoff, & Rosomoff, 2007). Along with the potency and duration of prescribed dosage there are also other factors entailing a patient's clinical history such as socio-economic and demographic characteristics, which together potentially lead an opioid naïve patient towards the increased risk of developing opioid use disorder. Given variation in patients' characteristics, there might be different paths that underline a patient's tendency or likelihood of developing opioid use disorder at a later stage after they have initiated opioid prescriptions. We investigate such paths as constituting distinctive patient-level characteristics. Identifying the influential factors associated with patients who at some point abuse these drugs is critical to determine the risk of future incidence of opioid use disorder. We hypothesized that there exists similarity in these factors for patients with opioid use disorder. Thus, our proposed framework starts with determining the influential features associated with opioid use disorder followed by the implementation and validation of machine learning algorithms as depicted in Figure 1 and Figure 2, respectively.

[Figure 1 near here]

[Figure 2 near here]

The features in the analytic file are ascertained before an opioid naïve patient develops or shows any sign of opioid use disorder in the follow up time period. Opioid naïve patients are identified from the pharmacy claims dataset. These are the patients who filed at least one claim that was identified as an opioid-related claim (identified using national drug codes) and had no other opioid prescriptions one year prior to the index date. The index date is chosen as the most recent fill date of an opioid prescription during the study time frame. Consistent with the prior studies (Ciesielski et al., 2016; Rice et al., 2012; White et al., 2009), for each opioid naïve patient, we take the first opioid prescription and from the date of that prescription we track the patient from 6 to 12 months further in time in the medical claims dataset to determine whether they had developed any sort of opioid use disorder. In order to identify such incidence of opioid use disorder, we used ICD-9 diagnosis codes associated with dependence and abuse of opioids and other drugs including heroin and methadone (see Table 5 in Appendix for corresponding clinical diagnosis and  associated ICD-9 codes). If an opioid naïve patient is diagnosed with any of those ICD-9 codes, we consider that patient as identified with opioid use disorder and without opioid use disorder otherwise. The first 6 months since the index prescription are used to collect information about the clinical diagnosis and opioid usage-pattern that had taken place once an opioid-naïve patient initiated opioid prescription. We use the term OUD to refer to patients with opioid use disorder and NOUD for patients without opioid use disorder. For both OUD and NOUD patients, we go one year back in time and gather their clinical history. Tracking of patients is censored when any incidence of opioid use disorder is identified within the follow up period. The outline of this conceptual framework is presented in Figure 3.

[Figure 3 near here]

### 2.3. Data Pre-processing, Feature Elimination and Implementation of predictive Model

Given that the success of a prediction model depends on how the features are being presented, we transform the raw MA APCD into influential features. MA APCD is a claim-level dataset with each record representing a unique medical or pharmacy claim associated with one patient. This also means that one patient can have multiple claims. This structure of dataset is not suitable for training machine learning models since it requires a patient-level instead of a claim-level dataset. Therefore, we aggregate claim-level information into patient-level information to make the dataset suitable for training and testing machine learning models. This includes capturing patient's demographic as well as clinical history as an independent feature of the dataset that may or may not have significant association with opioid use disorder.

This process entails following steps. The outline of these steps is presented in Figure 4, and the details are given in the following subsections.

### 2.3.1. Dealing with Missing Values and Feature Engineering

Age (categorical): Age for approximately 0.72% patients without opioid use disorder were missing, which we impute with mean age of all other patients without opioid use disorder (after considering overall age distribution). We consider age as categorical feature instead of a continuous one. Age is divided into five different buckets: i) 18-25, ii) 26-35, iii) 36-55, iv) 56-64, and v) above 65 in an effort to understand whether or not a patient belonging to a certain age-bin is more likely to develop opioid use disorder than other.

Gender (categorical): Approximately 0.02% records had gender values either missing or 'unknown'. The distribution of remaining data for male/female is approximately 43:57, so instead of imputing missing or unknown values as either male or female, we remove those records.

Degree of chronicity (categorical) of opioid usage: We engineer this feature to define the chronicity of opioid usage of opioid-naïve users. We use Proportion of Days Covered (PDC) to determine such chronicity of opioid-usage. PDC is defined as the fraction of days a patient was on opioid medications within one year since their first opioid prescription. We then categorize the PDC value into four levels. Patients with a PDC less than 20% are defined as non-chronic, whereas those with a PDC value greater than 80% are defined as highly chronic users of opioids. A PDC value within the range of 20% to 49%, and 50% to 79% are attributed to less chronic and moderately chronic users, respectively.

ICD-codes (categorical): In the data, the primary diagnoses were recorded as standard ICD-codes (which are alpha-numeric characters). As these ICD codes are difficult to interpret in their original form, we replace ICD-codes with their actual descriptions to make the data and results more interpretable.

[Figure 4 near here]

*2.3.2. Eliminating Features with Low Variance*

The final analytic file comprises approximately 600,000 patients and 12,000 features. Apart from age, sex, and zip code, all other features entail patients' clinical history extracted as ICD-9 codes. If a patient is diagnosed with a particular ICD-9 code, then we record that as n or 0 (where n is the number of times that patient is diagnosed for that particular ICD-9 code). This dimension comprising 600,000 patients and 12,000 features made further analysis difficult, so we eliminate less important features. When an analytic sample entails a large number of features (i.e., high dimensional), the training time for the model grows exponentially, and the risk of model overfitting also increases with number of features. The feature selection methods such as the one like Variance Threshold (VT) helps address these issues by reducing the number of features without losing much of the total information. Without training any model, we are required to eliminate features, so we chose the variance threshold (VT) method, which removes features with low variance. This algorithm only looks at the features (or predictors) and not the outcomes (or target). So, it is similar to an unsupervised machine learning technique. It removes features that have the same value for all the records in the sample. By having multiple features in the dataset, we expect to capture certain variance in the data that can help us in predicting the outcome. We use the VT method to calculate the variance of each feature and then only retained features with variance higher than the pre-defined threshold. Before adjusting a threshold for VT, we examine the distribution of variances of all the features and pick a threshold equal to 0.03 that covers most of the features, which capture some level of variance in the dataset. Meaning, if variance of a feature is close to 0, it is expected that most of its values are very close to each other (or same) and it does not help in distinguishing one patient from the other, thus does not contribute significantly when we predict opioid use disorder. There exists some general guidelines as per scikit-learn's documentation (scikit_learn, 2019): Suppose that we have a dataset with Boolean features and we want to remove all features that are either one or zero (on or off) in more than 80% of the samples. Boolean features are Bernoulli random variables, and the variance of such variables is given by $Var(x) = p(1 - p)$.

### 2.3.3. Retaining Most Important Features

After VT, we had 3076 features with a variance $\geq 0.03$. Our goal is to make the models more interpretable and computationally more efficient. Therefore, we also implement Chi$^2$ test, following VT, to retain only those features that are statistically significant with respect to target (i.e., opioid use disorder). A two tailed statistical significance test with a p-value set at $< 0.05$ is performed, reducing to 2628 features, which are used to train the predictive models.

To further reduce the number of features and improve models' interpretability, we use recursive feature elimination (RFE) technique, which removes (or eliminates) features recursively by pruning the original set of features from the model. The RFE takes a model trained with all the available features and assigns weights to whole set of features (either as coefficients for linear model or feature importance for tree-based model). In each iteration, the least important features are pruned based on weights. By doing this, it achieves the goal of getting a smaller set of features in each iteration and eventually ending up with only those features that have the most predictive power. As such, we are required to decide how many features we want to prune at the end of each iteration and provide a stopping condition for the RFE to terminate in order to get a feasible result. Using cross-validation technique, we set to remove 10% of the features after each iteration. The reason behind choosing 10% as the fraction of features to be removed after each iteration is to reduce the computational burden while still performing as many iterations as feasible. To develop a clinically interpretable model, we need to determine the set of features, which can provide higher area under receiving operating characteristics curve (AUC) and eliminate redundant features from the model. Reducing number of features (i.e., from 2628) by recursively training four different models and investigating their predictive performance on the test data was computationally expensive. Therefore, in every iteration, we have chosen 10% of the remaining features to be eliminated from the analytic sample. However, this 10% can be considered as a sensitivity-parameter of our framework which can be tuned to assess its impact. In order to set a stopping conditionfor RFE, we further investigate the change in AUC value by reducing the number of features from 2628, and determine the number of features that provide the maximum AUC values for different predictive models.

Another subtle consideration with RFE is that if we use it for different types of models such as Logistic Regression, Decision Tree, Random Forest etc., then it produces (or retains) different sets of features for different models when the algorithm terminates. For example, if we want to retain 50 features and we use RFE with logistic regression and random forest, we may end up with two different sets of 50 features from these two models. That is, these 50 features may or may not be same for two different models.

*2.3.4. Handling Class Imbalance Data*

In the study sample, the prevalence of patients identified with opioid use disorder is approximately 1%, which represent a class imbalanced sample. Therefore, selection of an appropriate performance metric is critical to evaluate the models' performance. To achieve better results, we implement Synthetic Minority Oversampling Technique (SMOTE)—a technique to over-sample a minority class (OUD patients). The SMOTE generate 'synthetic' samples of a minority class (OUD patients), resulting a similar distribution of NOUD and OUD patients in the dataset.

### 2.4. Implementation of Machine Learning Algorithms to Facilitate the Proposed Risk Prediction Model

We investigate and evaluate the performance of several well-known machine learning algorithms to precisely classify the individuals at risk for opioid use disorder. We frame this problem as a binary classification problem *(i.e., target-variable is: '0' if patient is identified with* opioid use disorder *and '1' if identified without* opioid use disorder). As we did not have labeled data, i.e., we did not have a *target-variable* that would tell us if a person had opioid use disorder, we were required to infer the labels using clinical history associated with opioid-dependence, abuse or overdose from the medical claims file.

Looking into the final analytic file, we observed big sparse features along with high-dimensional and non-linear feature space that may cause Logistic Regression based traditional classifiers to fail in terms of learning features or overfitting issues. We intend to adopt multiple predictive modeling approaches to present an analytic pipeline for predicting risk of opioid use disorder. More specifically, we intend to assess the performance of traditional Logistic Regression technique and also tree-based classifiers (Decision Tree, Random Forest, and Gradient Boosting).

At first, we split the analytic file into train-test set (70/30 split). As the dataset is highly imbalanced (i.e., prevalence of OUD class is ~1%), we perform SMOTE to balance the training data and get a 50/50 distribution of NOUD and OUD class. As a next step, we train different models as mentioned before. After training each model on an entire training dataset, we evaluate performance on both train and test data by using precision, recall, F1-score, and AUC value. We further eliminate features via RFE technique as discussed before and end up with different set of features for different models that contain the most amount of predictive power. Finally, we train the model again with remaining most important features and evaluate performance as explained earlier.

## 3.    Results and Discussions

The study sample consists of 44.51% male, of which 2.25% are identified with some aspects of opioid use disorder. The rest of the patients (55.49%) are female, and among them 0.81% had at least on diagnosis attributed to opioid use disorder. Out of approximately 66% of total patients with age above 36, a significant proportion (35.61%) had age in between 36 and 55, whereas only 18.21% and 12.2% had age within 56 to 65 and above 65, respectively. The proportion of patients with age below 36 is approximately 33%, among them 11.75% had age within 18 to 25 and 15.28% were with age 26 to 35. Relatively, a small proportion of patients (6.94%) were found with age less than 18 in the study sample. The majority of the study sample with opioid use disorder had age in between 18 to 35, whereas a relatively lower number of patients with opioid use disorder are identified in the age group 36 to 55. Such proportions of patients with diagnosed opioid use disorder were much less for patients with age below 18 and above 55.

### 3.1.  Machine Learning Approaches and Comparison of Predictive Performance

As mentioned in the previous section, the prevalence of patients with opioid use disorder is only ~1%. Therefore, accuracy alone cannot be considered as a reliable performance measure and choosing an appropriate evaluation metric is critical for selecting the best predictive model. As such, for each model, we determined precision, recall, F1-score and AUC, which is presented in Table 1 and Table 2, and graphical comparison is shown in Appendix (Figure 9 and Figure 10). We chose recall and AUC considering the following two reasons:

i.    We want to identify as many patients with opioid use disorder as we can. Identifying an OUD patient is more critical than misclassifying an NOUD patient, because if an OUD patient is misclassified, then it is likely that this patient will go without proper medication management and become dependent on opioids, potentially leading to the increased risk of drug overdose. On the other hand, if we misclassify a NOUD patient as an OUD patient then at most the medical practitioner will be more careful and may end up with prescribing pain medicine with lower dosage and for short duration of time, which will likely to be even less addictive. Although such NOUD patients are at relatively less elevated risk of developing opioid use disorder, due to the misclassification as OUD, the prescriber will have to closely monitor those patients to properly taper the dosage and reduce the harms of opioid withdrawal symptoms. Additionally, the prescriber will also need to be aware of the physical dependence on opioids, which is different from addiction and can result even when a patient properly takes them as prescribed.

ii. AUC is an effective indicator of model's ability to distinguish a rare class from the prevalent one.

Based on these two performance measures, we present a comparative analysis across all four predictive models. Figure 7 shows the area under the receiving operating characteristic (ROC) curves for all four models. The value of this metric varies between 0 and 1, and the higher the values the better the model's performance. In other words, we expect the curve to reach top-left corner and thus have a value close to 1. The AUC in case of test data are highest for Random Forest and Gradient Boosting for any value of false positive and true positive rates, and it is slightly lower for Decision Tree model. It is clear that the AUCs for tree-based models significantly outperform the AUC of the Logistic Regression model.

[Table 1 near here]
[Table 2 near here]

After performing VT and chi-squared test on training sample, the recall value while predicting OUD patients for Logistic Regression, Decision Tree, Random Forest, and Gradient Boosting were 0.95, 0.99, 0.99, and 1.0, respectively (Table 1). And, the AUC value in above-mentioned order of four algorithms were 0.965, 0.989, 0.985, and 0.994, respectively (Table 1). However, the performance of these models is not equally well in case of test sample; both recall, and AUC dropped significantly, apparently indicating models' overfitting issue. Then, while eliminating redundant features using RFE, with 49 features Random Forest model achieved an AUC value of 0.990 and 0.970 for training and test set, respectively as presented in Figure 5. The results from similar analyses for Gradient Boosting, Decision Tree, and Logistic Regression model are provided in Figure 13, Figure 14, and Figure 15, respectively as presented in Appendix. Both the Gradient Boosting and Logistic Regression models achieved maximum AUC value with 49 Features, while Maximum AUC for Decision Tree model was observed with 13 features.

[Figure 5 near here]

Moreover, after RFE, the recall of Random Forest model in predicting OUD also significantly improved to 0.97, which is higher than 0.88 as found after VT and chi-squared test in case of test sample. Such recall of Random Forest model is also higher than what observed for Decision Tree (recall 0.92), Gradient-Boosting (recall 0.92), and Logistic Regression (recall 0.93) after RFE in case of test set. Besides, in terms of AUC, Random Forest model (AUC = 0.970) outperformed Decision Tree (AUC = 0.956), Gradient-Boosting (AUC = 0.956), and Logistic Regression (AUC = 0.935) as evident in Table 2 and Figure 7. Figure 6 presents the comparison of model's recall after VT in combination with Chi-squared test, and RFE.

[Figure 6 near here]

[Figure 7 near here]

We further use Chi-square statistic to rank the final models (after RFE) based on features' p-values from $Chi^2$ test. Since the most important features for each model are retained after RFE, all of those features are statistically significant based on Chi-squared test (i.e., p-value < 0.05). We, then calculate the average of p-values for all the features in each model and find that Random Forest and Gradient Boosting models have almost same average p-values for their features. However, the same for Logistic Regression is turned out to be much higher, indicating the fact that the final feature set of Logistic Regression is less statistically significant than that of Random Forest and Gradient Boosting. Moreover, the average of p-values for all the features of Decision Tree is close to 0, indicating that the features are most statistically significant with respect to target. However, we have observed that the Random Forest model provides more accurate predictions than Decision Tree in terms of AUC and recall. This could be because of the way the two models work. Since, it is important to accurately identify OUD patients, we still consider Random Forest as best model based on our analysis.

Some meaningful insights from Logistic Regression model is obtained based on odds ratio (OR) associated with most important features. For this model, we consider female, age above 65, and non-chronic use of opioids as reference group for categorical variables with multiple levels. The top 10 features that have the highest $OR$ are: moderate (13.66) and high chronic use of opioids (12.25), poisoning by heroin (12.79), age 26 to 35 (11.27), age 18 to 25 (7.47),age 36 to 55 (4.13), male (1.81), poisoning by unspecified opium (alkaloids) (4.32) or opiates and narcotics (5.63), and less chronic use of opioids (4.14). Patients attributed to high chronic use of opioids are approximately 12.25 times more likely to develop opioid use disorder than did non-chronic use of opioids. Similar comparison can be made for patients with moderate and less chronic use of opioids. Compared to individuals with age above 65, patients aged in between 26 and 35 are 11.27 times more likely to experience opioid use disorder. Similarly, male patients are 1.81 times more likely to have diagnosed with opioid use disorder than female. The rest of the top 10 features are discrete. $OR$ for poisoning by heroin is 12.79, which means that if a patient is diagnosed once for poisoning by heroin, then he/she is approximately 12 times more likely to develop opioid use disorder as compared to patient who did not have diagnosed with heroin poisoning.

From the feature importance plot of reduced Random Forest model (Figure 8), we see that along with previous diagnosis of opioid dependence, degree of chronicity, and age 26 to 35 are very important predictors of opioid use disorder. Such feature importance plots of Decision Tree and Gradient Boosting model are presented in Figure 11 and Figure 12 in Appendix. A brief clinical description of the top features obtained from all the four predictive models after RFE are presented in Table 3 (see Appendix) along with the corresponding ICD-9 codes. The important features in the Decision Tree model after RFE are somewhat similar to those we obtian from Random Forest model. In case of Gradient Boosting model, we see that this model gives less importance to age-related features as compared to Random Forest and Logistic Regression models.

[Figure 8 near here]

### 3.2. Discussion

This study investigates the risk factors that lead an opioid naïve user towards the increased risk of opioid use disorder, and also leverage several machine leaning algorithms to predict the likelihood of such disorder. Unlike existing studies on predicting opioid dependency and abuse, we allow the selected predictive model to ascertain the risk factors associated with opioid use disorder rather than building a model using pre-selected features. Often times, it is argued that administrative claims datasets lack adequate clinical details for developing efficient and reliable predictive models for patient. However, our study demonstrates that such an approach is feasible and clinically significant in predicting potential opioid use disorder by extracting influential clinical factors using large-scale healthcare administrative claims data.

Based on the comparison of several performance measures, our study shows that tree-based models, in general, outperform a Logistic Regression model, and among the tree-based models, Random Forest superiorly performs in predicting opioid use disorder with clinically significant features and higher predictive accuracy, potentially making this model applicable in clinical setting. The most important demographic features includes patients' age between 18 to 25 and 26 to 35 , and male gender, a findings consistent with (Dufour et al., 2014), which reported increased risk of opioid use disorder for male patients. Among clinical factors, our model determined that prior history of opioid dependence, poisoning by heroin, opioid abuse, and prior evidences of drug withdrawal adverse effect related to mental illness or psychotic disorder, abuse of non-opioid drugs or substances, poisoning by unspecified drugs or medical substances, and dependence on unspecified alcoholic substance are strong predictors of opioid use disorder which supports several prior research (Edlund et al., 2010; Rice et al., 2012). Other clinical features entail previous history of dependence on unspecified drugs and (or) combinations of opioid type drugs with other drugs, anxiety, depressive disorder, poisoning by opium (alkaloids), and long-term use of other medications add significantly valuable knowledge to what is already known regarding the risk factors of opioid use disorder.

Because the prior history of opioid dependency has already been proven as a significant predictor of developing future dependence on opioid, it may be argued that inclusion of this feature may result in limited clinical significance of the predictive model. Towards a deeper investigation on this issue, we further train the Random Forest model excluding all the features that are related to any kind of prior history of opioid dependency and abuse. The reason behind considering the Random Forest model in this regard is its superiority over all other three predictive models. Results (presented in Table 4 in Appendix) demonstrate that exclusion of those features reduced the model AUC value to 0.833. In addition to that, we also observe a much lower value for model recall, implying that the model significantly failed to predict those OUD patients who actually had opioid use disorder. Therefore, such a predictive model will have a greater chance to potentially mislead the physicians, leading to enhanced risk for patients who are more likely to develop opioid use disorder.

While several of our findings are noticeably consistent with prior clinical research examining the association of risk factors with different aspects of opioid use disorder (Dufour et al., 2014; Ives et al., 2006; Rice et al., 2012; White et al., 2009), the predictive power of our analysis substantially adds value to the existing literature. Prior research also included the history of substance abuse (Ciesielski et al., 2016; Dufour et al., 2014; Glanz et al., 2018; White et al., 2009) and opioid abuse (Edlund et al., 2010; Rice et al., 2012) as predictors in the logistic regression model. However, those studies did not apply machine learning techniques and achieved suboptimal predictive performance as indicated by a lower value of c-statistic (alternative measure of AUC) within a range of 0.75 to 0.92 (Ciesielski et al., 2016; Dufour et al., 2014; Glanz et al., 2018; Rice et al., 2012; White et al., 2009). Moreover, the accuracy reported by c-statistic in the existing studies (Dufour et al., 2014; Rice et al., 2012; White et al., 2009) might also be overestimated given the fact that the dataset was highly imbalanced with a reasonably smaller proportion of total study sample of patients identified as opioid abusers or dependent. Such imbalance issue, if not overcome properly could lead to an overfitted predictive model and biased estimation of accuracy value, which could potentially question the applicability and reliability of predictive model in such a highly critical application area. Our study, with the help of SMOTE takes care of this issue and achieves higher Recall and AUC value, which overcome the limitation of biased learning from imbalanced dataset.

Although, one study (Lo-Ciganic et al., 2019) that used machine learning to predict risk of opioid overdose considered patients' prior history of prescription opioid overdose, substance, and opioid use disorder and achieved a c-statistic of 0.90 for a Gradient Boosting model, it was limited within Medicare individuals aged above 65. Besides Logistic Regression technique, we also implement other tree-based algorithms and obtain an AUC value of 0.97. Because our analytic sample entails a large number of features with sparsity and skewness of features, it is perhaps not surprising that we found that tree-based models outperformed a Logistic Regression model in terms of AUC and recall value. As Logistic Regression belongs to a family of Generalized Linear Models (GLM), it can capture the linear relationship between predictors and outcomes, effectively. However, it is likely to fail in capturing some complex relationships. Because tree-based models are just structured hierarchy of rules to make predictions (not necessarily linear), they are robust to some above-mentioned data issues, and thus they outperform the Logistic Regression model. As such, we suggest that researchers studying healthcare claims data to predict opioid use disorder could consider evaluating and implementing more than one classifier. Particularly, we recommend using the Random Forest model as it is evident from our analysis that the Random Forest model is able to identify OUD patients with higher accuracy as compared to a Logistic Regression model while also efficiently determined the risk factors that are significant predictors of opioid use disorder. Hence, our investigation provides meaningful insights on risk factors associated with opioid use disorder along with a predictive decision support system with higher reliability that can be instrumental for healthcare practitioners.

## 4. Conclusions, Limitations and Future Research Directions

This study provides a machine learning framework that intends to serve as a decision support system for physicians at the point of care or initial opioid prescription to help identify the patients at risk of future incidence of opioid use disorder. To the best of our knowledge, this is the first study utilizing MA APCD to develop a predictive analytics framework, particularly focusing on commercially insured prescription opioid users. While prior research largely dependent on the Logistic Regression based predictive technique suffered from overestimated accuracy due to imbalance data, our study tackles this issue utilizing SMOTE, and proposes a set of tree-based predictive models that achieved superior predictive performance over Logistic Regression model. We further utilize VT, Chi-squared test, and RFE techniques in an effort to enhance model computational efficiency and interpretability. Moreover, we are able to show that along with higher predictive performance, Random Forest model extracts the risk factors from patients' one-year clinical history, some of those are consistent with prior findings of clinical literature. Thus, our study enables researchers and healthcare decision makers to utilize machine learning approach on healthcare claims datasets and provide with the ability to predict opioid use disorder.

We note several limitations of our study. It was not possible to track patients' clinical history before 2011 due to the unavailability of the data. Because data were not available beyond 2015, we were not able to investigate how many of them were diagnosed with opioid overdose in the future. As such, we had to exclude years 2011 and 2015 from being considered as an index year.  In addition, we could not identify the patients with opioid use disorder in year 2014, as pursuant to federal policy, CHIA was required to remove all medical claims for drug dependence after 2013. Another limitation has to do with the selection of four (although well-known) out of the many other predictive algorithms. Besides, we have not implemented the predictive analytics framework in clinical setting, nor assessed the actual impact of such implementation on clinical decision, and we acknowledge that these are the issues with significant importance. However, we have successfully demonstrated that modern machine learning techniques can effectively and feasibly be used on large-scale healthcare claims datasets, particularly in predicting the risk of opioid use disorder for commercially insured individuals. Although SMOTE has reduced the bias caused by the majority class, it may perhaps result some overfitting. The synthetic observation created by SMOTE are usually sensitive to the specific samples of minority class that are selected form the true data distribution as SMOTE utilizes k-nearest neighbors (KNN) to create synthetic observations. Because, the nature of the true data distribution (e.g., whether or not it is uniformly distributed) is not known and it is possible that the minority class samples (used to create synthetic observations) can be unlikely due to the presumable non-uniform true distribution of the data, then there are possibilities to fit the model to observations not truly representative of the original data distribution. This may perhaps introduce some degree of overfitting and lack of generalization ability of the predictive model. Despite this, because we used SMOTE on the training dataset and not on the entire dataset prior to cross validation, the model prediction was performed on the unseen data-fold in the first pass of cross validation, and similar process was followed for rest of the data-folds. Therefore, the above-mentioned overfitting due to SMOTE did not have substantial impact during the training phase of the models.

Future study will extend implementing the current framework on claims dataset coming from other states that are highly afflicted with opioid overdose epidemic. The sensitivity of the proposed framework can also be investigated on other study sample including vulnerable population like Medicaid, and Veterans to reveal novel risk factors associated with the risk of opioid use disorder. Such analysis will also take into account other predictive algorithms to investigate and compare their predictive power with current findings. We also aim to develop a user interface so that physicians can utilize this as ready-to-use tool, which will generate a risk score for an individual before prescribing opioids once the required information associated with risk factors are provided.

**Funding**

This research was supported by the Global Resilience Institute (GRI), Northeastern University, Boston, USA

**Declaration of Interest Statement**

The authors confirm that they do not have any conflict (financial or other) of interest in regard to this study

**Appendix**

[Figure 9 near here]

[Figure 10 near here]

[Figure 11 near here]

[Figure 12 near here]

[Figure 13 near here]

[Figure 14 near here]

[Figure 15 near here]

[Table 3 near here]

[Table 4 near here]

[Table 5 near here]

**List of figures**

**Figures**

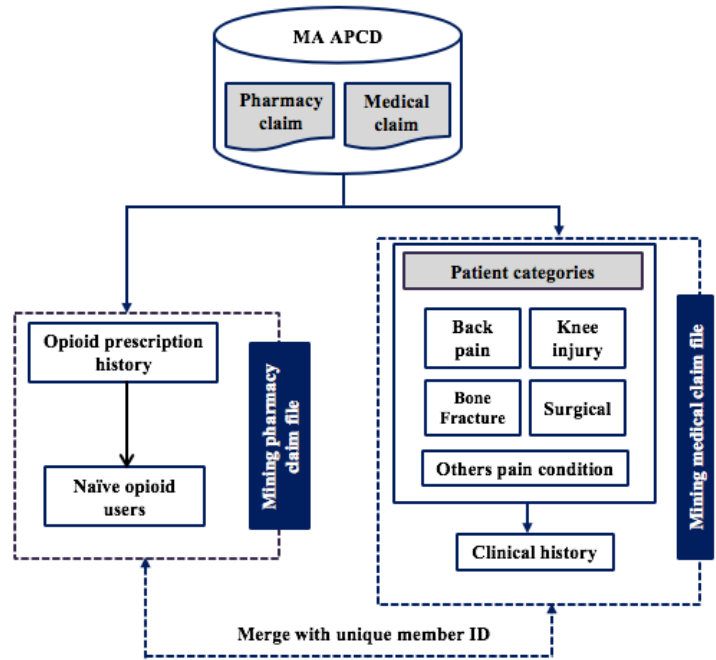

Figure 1. Outlining the development of analytic file from the MA APCD

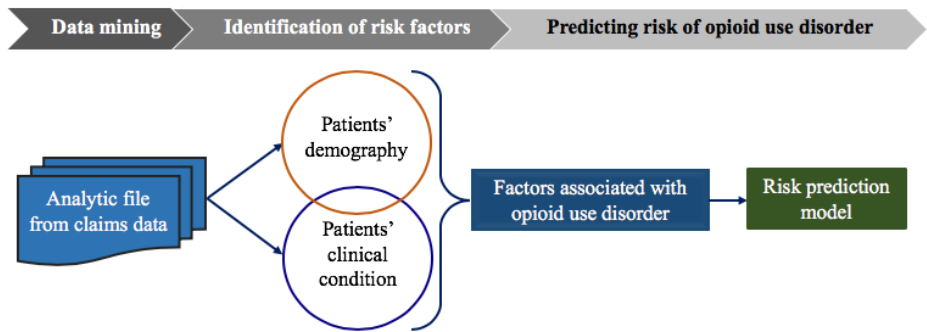

Figure 2. Outline of developing a predictive model for opioid use disorder

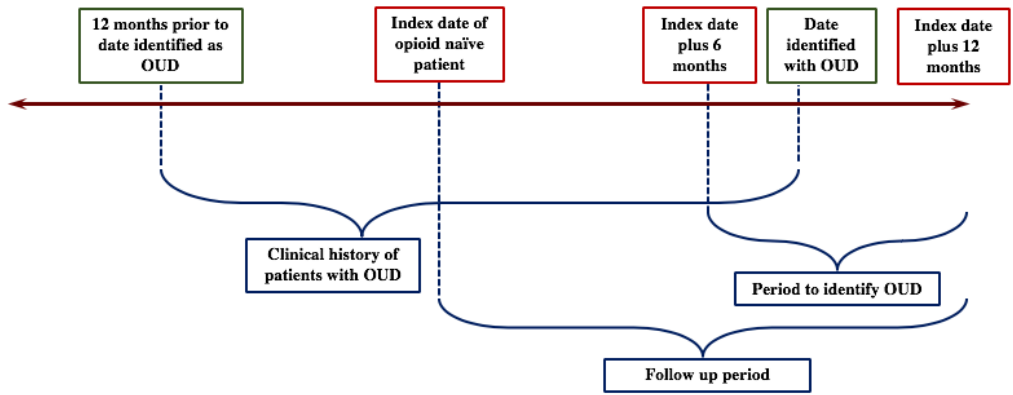

Figure 3. Conceptual framework for study design

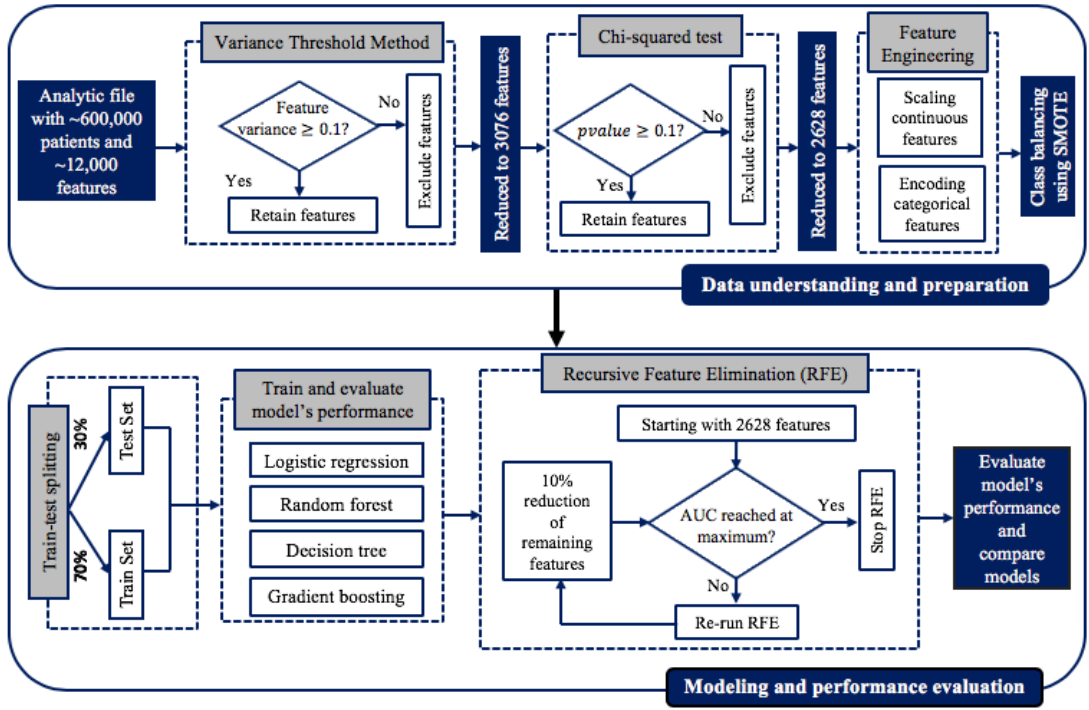

Figure 4. Data preparation, development and evaluation of predictive model

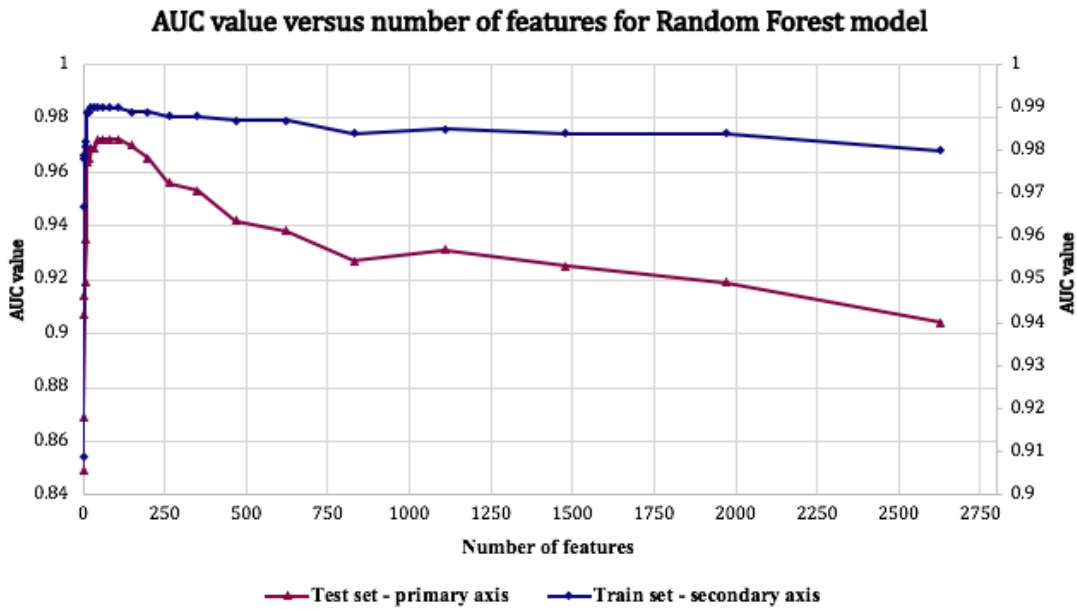

Figure 5. AUC value of Random Forest model after different iteration in RFE

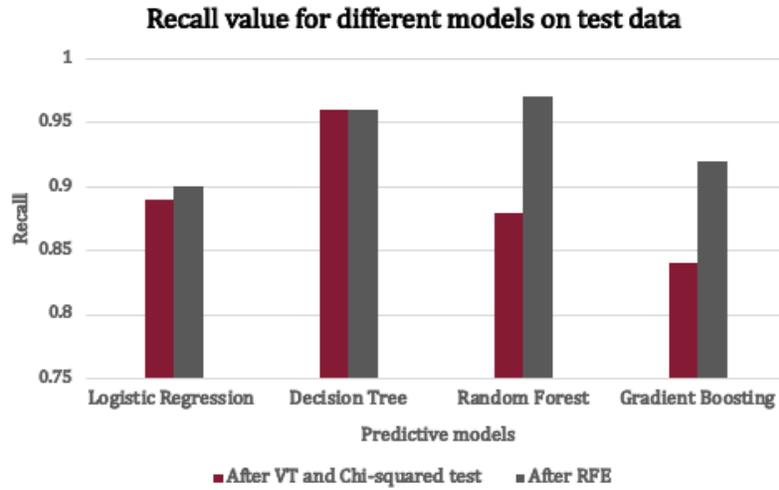

Figure 6. Comparison of recall value for different predictive models

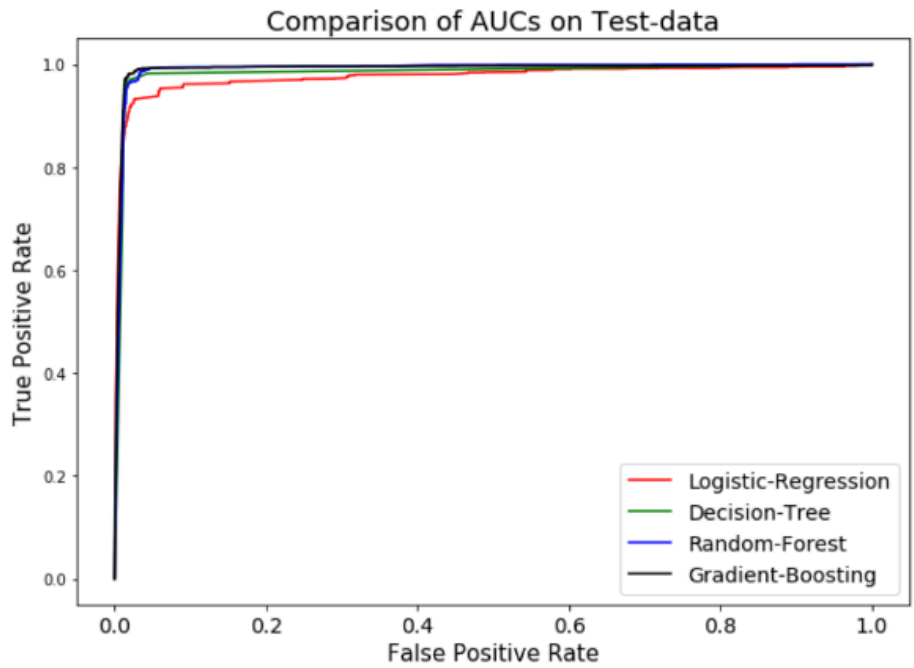

Figure 7. Comparison of ROC for different predictive models

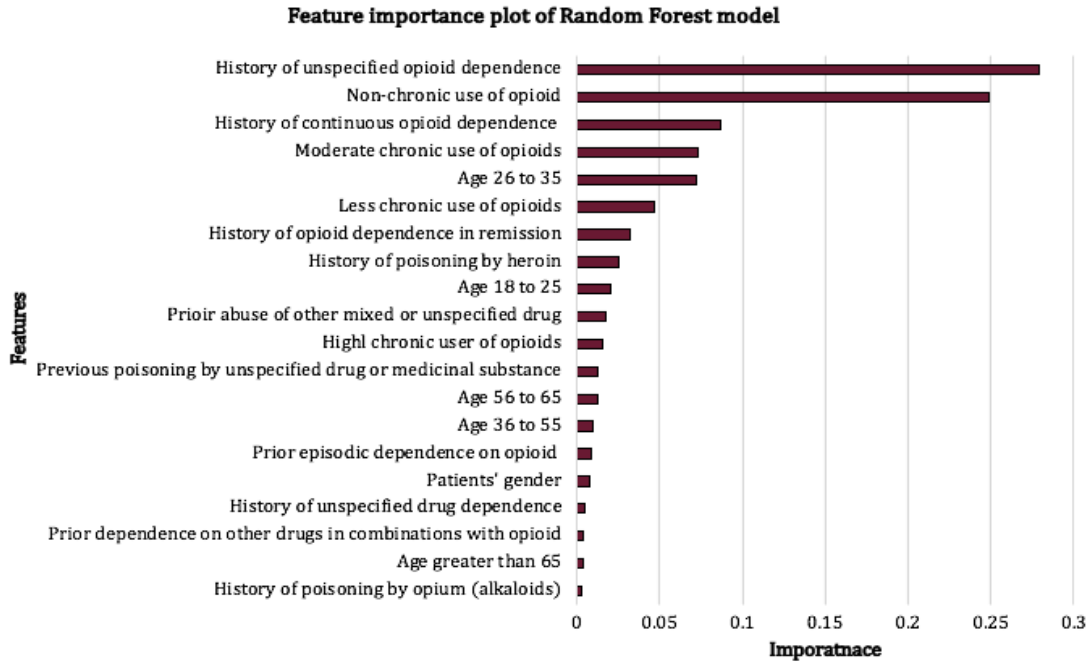

Figure 8. Feature importance plot for Random Forest model

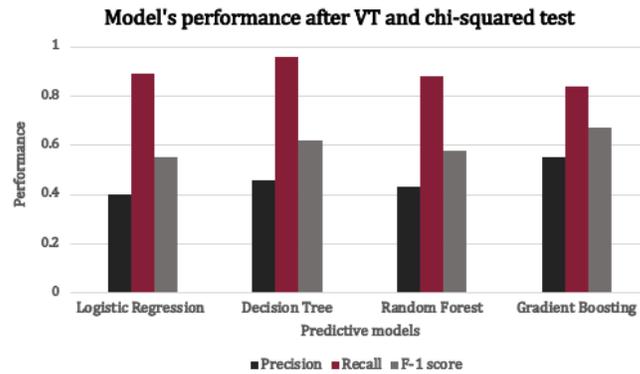

Figure 9. Comparison of models after VT and Chi-squared feature elimination methods

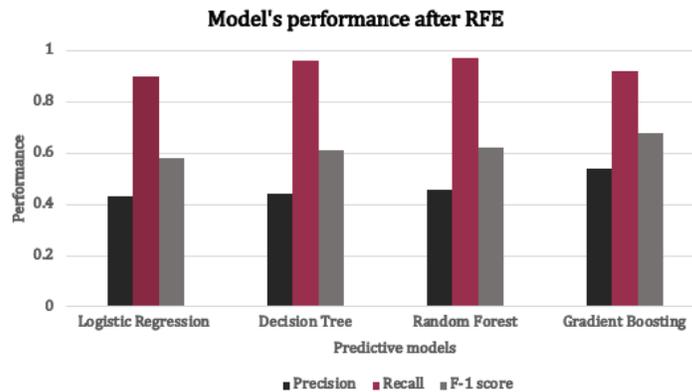

Figure 10. Comparison of models after RFE

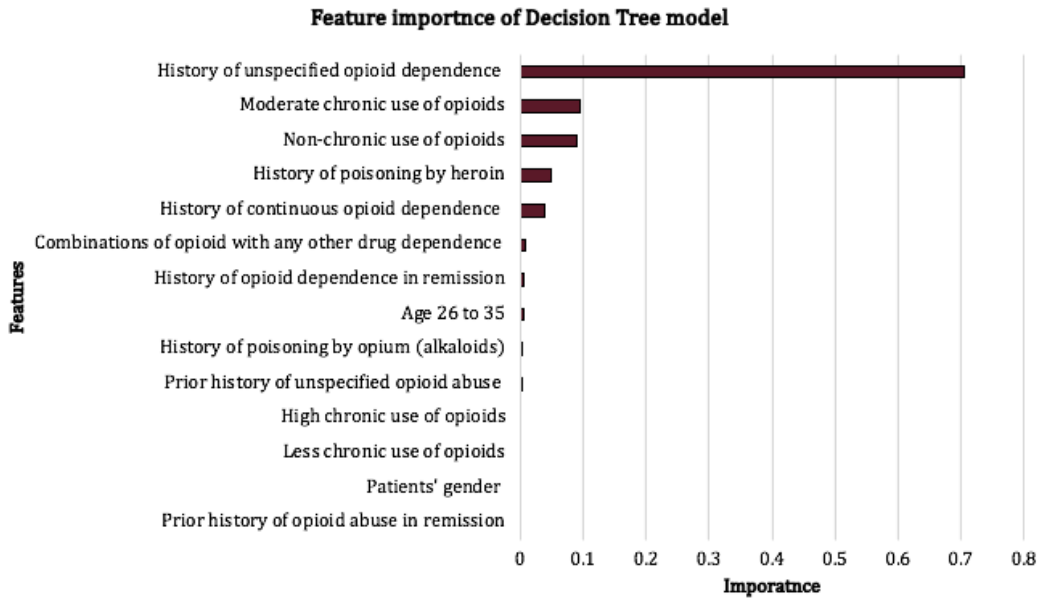

Figure 11. Feature importance plot for Decision Tree model

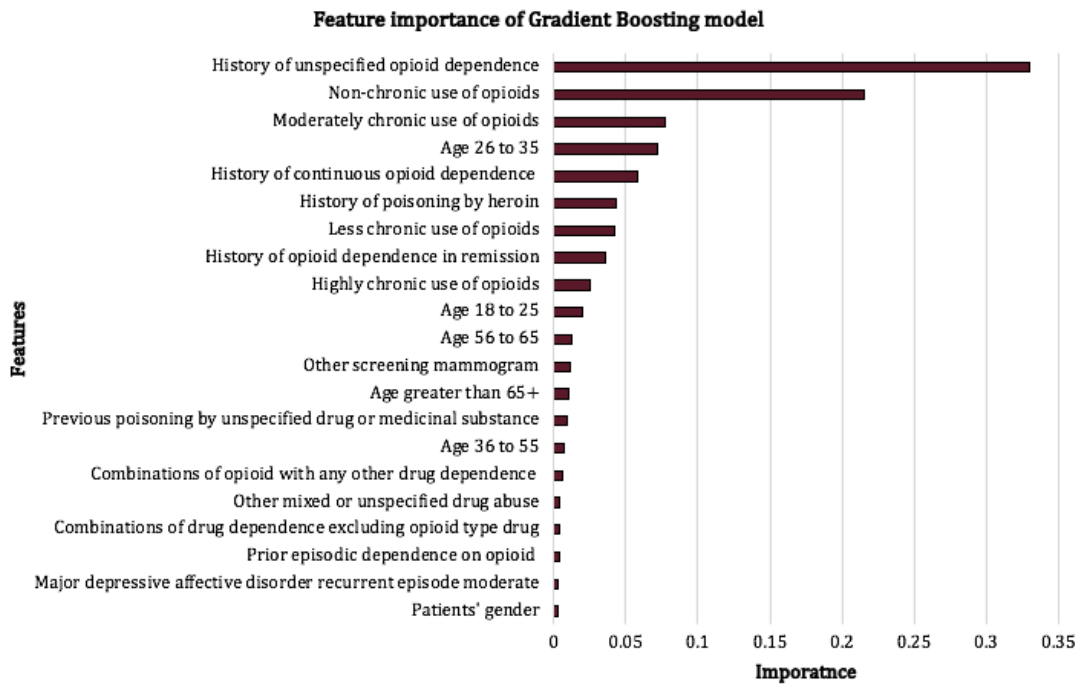

Figure 12. Feature importance plot for Gradient Boosting model

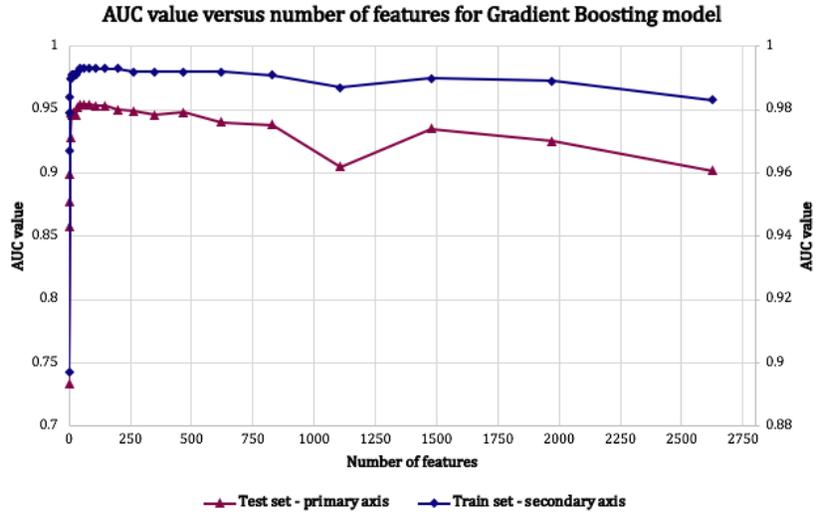

Figure 13. AUC value of Gradient Boosting model after different iteration in RFE

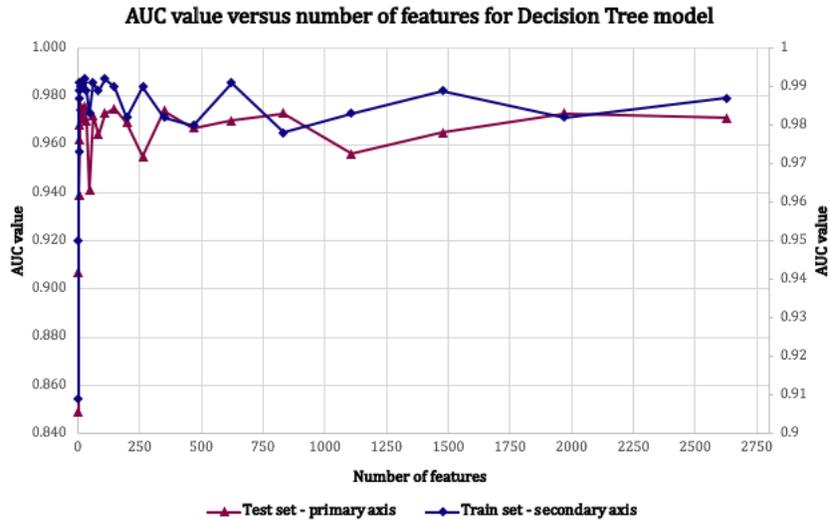

Figure 14. AUC value of Decision Tree model after different iteration in RFE

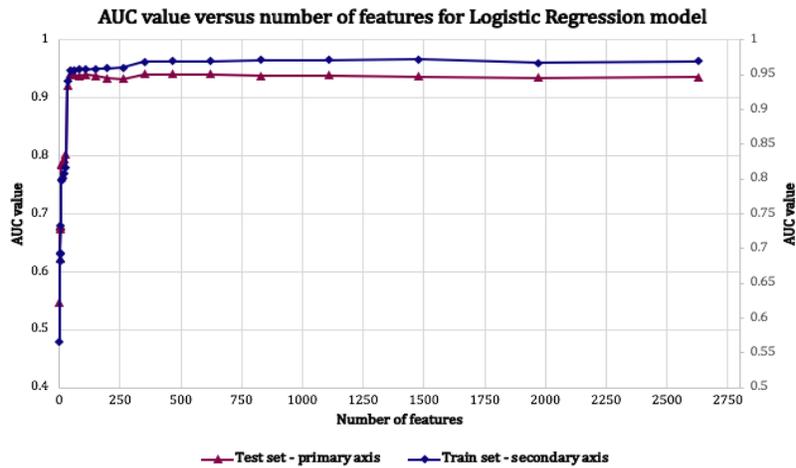

Figure 15. AUC value of Logistic Regression model after different iteration in RFE

**Tables**

**Table 3**
ICD codes of the clinical features obtained after RFE for four different models

| Features | Logistic Regression | Decision Tree | Gradient Boosting | Random Forest |
|---|:---:|:---:|:---:|:---:|
| History of poisoning by heroin | ✓ | ✓ | ✓ | ✓ |
| History of poisoning by opium (alkaloids) | ✓ | ✓ | ✓ | ✓ |
| Previous poisoning by other opiates and related narcotics | ✓ | | | |
| Prior dependence (continuous) on other drugs in combinations with opioid | ✓ | | ✓ | ✓ |
| History of opioid abuse in remission | ✓ | ✓ | ✓ | |
| Prior dependence (in remission) on other drugs in combinations with opioid | ✓ | | | |
| Prior episodic dependence on opioid | ✓ | | ✓ | ✓ |
| Previous poisoning by unspecified drug or medicinal substance | ✓ | | ✓ | ✓ |
| History of unspecified opioid abuse | ✓ | ✓ | ✓ | ✓ |
| History of episodic opioid abuse | ✓ | | | |
| History of continuous opioid dependence | ✓ | ✓ | ✓ | ✓ |
| History of unspecified opioid dependence | ✓ | ✓ | ✓ | ✓ |
| Prior dependence on other specified drug | ✓ | | | |
| Prior dependence (unspecified) on other drugs in combinations with opioid | ✓ | ✓ | ✓ | ✓ |
| Drug withdrawal | ✓ | | ✓ | ✓ |
| Prior dependence (episodic) on other drugs in combinations with opioid | ✓ | | | |
| History of opioid dependence in remission | ✓ | ✓ | ✓ | ✓ |
| Prior history of continuous opioid abuse | ✓ | | | ✓ |
| Chronic or unspecified gastric ulcer with hemorrhage, without mention of obstruction | ✓ | | | |
| Unspecified orthopedic aftercare | ✓ | | | |
| Other hemorrhagic disorder due to intrinsic circulating anticoagulants antibodies or inhibitors | ✓ | | | |
| Malignant neoplasm of trigone of urinary bladder | ✓ | | | |
| Lumbosacral root lesions not elsewhere classified | ✓ | | | |
| Impaired glucose tolerance test (oral) | ✓ | | | |
| acute myocardial infarction of unspecified site episode of care unspecified | ✓ | | | |
| Abnormal mammogram unspecified | ✓ | | | |
| Unspecified nonpsychotic mental disorder | ✓ | | | |
| Acute apical periodontitis of pulpal origin | ✓ | | | |
| Poisoning by other specified drugs and medicinal substances | ✓ | | | |
| Personal history of other malignant neoplasm of skin | ✓ | | | |
| Mitral valve insufficiency and aortic valve stenosis | ✓ | | | |
| Reflex sympathetic dystrophy of the upper limb | ✓ | | | |

| Features | Logistic Regression | Decision Tree | Gradient Boosting | Random Forest |
|---|---|---|---|---|
| Other and unspecified alcohol dependence | | | ✓ | ✓ |
| Generalized anxiety disorder | | | ✓ | ✓ |
| Unspecified viral hepatitis c without hepatic coma | | | ✓ | ✓ |
| Posttraumatic stress disorder | | | ✓ | |
| Cellulitis and abscess of upper arm and forearm | | | ✓ | |
| Senile nuclear sclerosis | | | ✓ | |
| Routine infant or child health check | | | ✓ | ✓ |
| Depressive disorder not elsewhere classified | | | | ✓ |
| Other alteration of consciousness | | | | ✓ |
| Encounter for therapeutic drug monitoring | | | | ✓ |
| Chronic hepatitis c without mention of hepatic coma | | | | ✓ |
| Laboratory examination unspecified | | | | ✓ |
| Tobacco use disorder | | | | ✓ |
| Examination of eyes and vision | | | | ✓ |

**Table 4**
Performance of the Random Forest model on the test dataset after RFE and excluding features related to prior history of opioid dependency and abuse

| Step | Class | Precision | Recall | F-1 score | AUC |
|---|---|---|---|---|---|
| Including opioid dependency features | NOUD | 1 | 0.98 | 0.99 | 0.97 |
| | OUD | 0.46 | 0.97 | 0.62 | |
| Excluding opioid dependency features | NOUD | 1 | 0.95 | 0.97 | 0.83 |
| | OUD | 0.19 | 0.72 | 0.3 | |

**Table 5**

List of diagnoses used to determine opioid use disorder

| Description of diagnoses used to determine opioid use disorder | ICD-9 codes |
| --- | --- |
| Opioid type dependence unspecified | 304.00 |
| Opioid type dependence continuous | 304.01 |
| Opioid type dependence episodic | 304.02 |
| Opioid type dependence in remission | 304.03 |
| Combinations of opioid type drug with any other drug dependence unspecified | 304.70 |
| Combinations of opioid type drug with any other drug dependence continuous | 304.71 |
| Combinations of opioid type drug with any other drug dependence episodic | 304.72 |
| Combinations of opioid type drug with any other drug dependence in remission | 304.73 |
| Opioid abuse unspecified | 305.50 |
| Opioid abuse unspecified | 305.51 |
| Opioid abuse episodic | 305.52 |
| Opioid abuse in remission | 305.53 |
| Poisoning by opium (alkaloids) unspecified | 965.00 |
| Poisoning by heroin | 965.01 |
| Poisoning by methadone | 965.02 |
| Poisoning by other opiates and related narcotics | 965.09 |
| Accidental poisoning by heroin | E850.0 |
| Accidental poisoning by other opiates and related narcotics | E850.2 |
| Heroin causing adverse effects in therapeutic use | E935.0 |
| Other opiates and related narcotics causing adverse effects in therapeutic use | E935.2 |